\documentstyle[aps,prb,preprint,psfig]{revtex}
\begin{document}
\title{Effect of external magnetic field on electron spin dephasing induced
by hyperfine interaction in quantum dots}
\author{Y. G. Semenov}
\address{Institute of Semiconductor Physics, National Academy of Sciences
of Ukraine\\
Prospekt Nauki, 45 Kiev 03028 Ukraine}
\author{K. W. Kim}
\address{Department of Electrical Computer Engineering\\
North Carolina State University, Raleigh, NC 27695-7911}
\date{\today }
\maketitle

\begin{abstract}
We investigate the influence of an external magnetic field on spin phase
relaxation of single electrons in semiconductor quantum dots induced by the
hyperfine interaction. The basic decay mechanism is attributed to the
dispersion of local effective nuclear fields over the ensemble of quantum
dots. The characteristics of electron spin dephasing is analyzed by taking
an average over the nuclear spin distribution. We find that the dephasing
rate can be estimated as a spin precession frequency caused primarily by the
mean value of the local nuclear magnetic field. Furthermore, it is shown
that the hyperfine interaction does not fully depolarize electron spin. The
loss of initial spin polarization during the dephasing process depends
strongly on the external magnetic field, leading to the possibility of
effective suppression of this mechanism.
\end{abstract}


%\begin{multicols}{2}

%\narrowtext
%%%%%%%%%%%%%%%%%%%%%%%%%%%%%%%%%%%%%%%%%%%%%%%%%%%%%%%%%%%
% BODY OF THE PAPER

The spin state of an electron confined in a semiconductor quantum dot (QD)
is considered one of the most promising candidates for realizing the basic
building block (i.e., qubit) of a quantum information system. Since the
fundamental concept of this new paradigm relies on quantum mechanical
entanglement of qubits, it is quite crucial to control spin relaxation
processes that destroy the coherence of spin quantum state. So far, most of
the attention has been devoted to the relaxation processes that result in
irreversible loss of wave function phase due to spin-phonon interaction
caused by spin-orbital coupling in solids or hyperfine interaction (HFI) in
crystals with non-zero nuclei spin moments (see Refs.\ \onlinecite{KhaetNaz} 
and \onlinecite{Erling} as well as the references therein). 
The common feature of these spin-lattice mechanisms is that their relaxation
rates are very small in QDs at low temperatures.

On the other hand, the HFI can be considered as a source of a local magnetic
field $\overrightarrow{H}_{HF}$ acting on the electron spin that does not
disappear at low (or even zero) temperature.~\cite{OptOr} This particularity
makes the HFI a potentially dominant mehchanism at sufficiently low 
temperatures. In typical
QDs, a sum of the contributions from a great number of nuclei spins forms
this field. Thus, the strength and the direction of the $\overrightarrow{H}%
_{HF}$ are the random variables, which vary from one QD to the next.
Obviously, this dispersion can be damaging to quantum computation since
electron spin precession occurs with a random phase and frequency.
Nevertheless, it appears from a qualitative speculation that the role of $%
\overrightarrow{H}_{HF}$ dispersion diminishes progressively with an
increasing strength of the homogeneous external magnetic field $%
\overrightarrow{B}$ applied to the array of QDs.

In this paper, we provide a quantitative analysis of electron spin evolution
under the presence of an external magnetic field $\overrightarrow{B}$ as
well as the local hyperfine field $\overrightarrow{H}_{HF}$. Note, that a
theory of electron spin relaxation caused by HFI in a QD was recently
presented in Ref.\ \onlinecite{KhaetLoss}. The main interest of Khaetskii {\em et
al.}~\cite{KhaetLoss} lies on the electron spin decoherence process inside a
single QD when the external magnetic field is zero. It also contains a brief
discussion of spin dephasing time. On the other hand, the present study
concentrates on the important process of electron spin dephasing induced by
HFI in an ensemble of QDs and explicitly considers the effect of external
magnetic fields.

The Hamiltonian of electron spin $S$ in a QD containing $N$ nuclear spins $%
I_{j}$ ($j=1,...,N$) takes the form ($\hbar =1$)%
\begin{equation}
{\cal H}=\omega _{e}S_{z}+\omega _{n}\sum_{j=1}^{N}I_{jz}+\sum_{j=1}^{N}A_{j}%
\overrightarrow{{S}}\cdot \stackrel{\rightarrow }{I}_{j},  \label{eq1}
\end{equation}%
where $\omega _{e}$ and $\omega _{n}$ are electron and nuclear spin
splitting in a magnetic field directed along the $Z$ axis, $A_{j}$ is a
constant of HFI with the $j$-th nuclear spin. The Hamiltonian of Eq.\ (\ref%
{eq1}) is isomorphic to that, which was introduced to describe the bound
magnetic polaron in diluted magnetic semiconductors. Most particularities of
optical spectroscopy as well as the thermodynamics of bound magnetic polaron
were successfully described by a model where differences in constants of
spin-spin interaction were ignored.~\cite{RybSem,Durst} By analogy with this
model, we describe the dynamics of electron spin by Hamiltonian with a
single effective constant $A$ of HFI. This allows one to express the
Hamiltonian in terms of the total nuclear spin moment $\stackrel{\rightarrow 
}{F}=\sum_{j=1}^{N}\stackrel{\rightarrow }{I}$: 
\begin{equation}
{\cal H}=\omega _{e}S_{Z}+\omega _{n}F_{z}+A\overrightarrow{{S}}\cdot 
\overrightarrow{F}.  \label{eq2}
\end{equation}%
Even with such a reduced Hamiltonian, the corresponding equations for the
motions of electron and nuclear spins are quite complex, 
\begin{equation}
\frac{d}{dt}\left\langle {\vec{S}}\right\rangle = \overrightarrow{B}%
_{e}\times \left\langle {\vec{S}}\right\rangle  +A\left\langle 
\overrightarrow{F}\times {\vec{S}} \right\rangle ;  \label{eq3}
\end{equation}%
\begin{equation}
\frac{d}{dt}\left\langle {\vec{F}}\right\rangle = \overrightarrow{B}%
_{n}\times \left\langle \overrightarrow{F}\right\rangle 
+A\left\langle  {\vec{S}\times }\overrightarrow{F}
\right\rangle .  \label{eq4}
\end{equation}%
where $\overrightarrow{B}_{e}=\left\{ 0,0,\omega _{e}\right\} $ and $%
\overrightarrow{B}_{n}=\left\{ 0,0,\omega _{n}\right\} $ are in units of
energy. Assuming that the flip-flop processes are unimportant for the
problem we consider, the $\Psi $-function of the Hamiltonian given in Eq.\ (%
\ref{eq2}) can be factorized with respect to electron and nuclear spins.
This means that $\left\langle  \overrightarrow{F}\times {\vec{S}}%
\right\rangle =\left[ \left\langle \overrightarrow{F}\right\rangle
\times \left\langle {\vec{S}}\right\rangle \right] $, and Eqs.\ (\ref{eq3})
and ({\ref{eq4}) present the closed system of equations with respect to $%
\left\langle {S}_{\alpha }\right\rangle $ and $\left\langle {F}_{\alpha
}\right\rangle $ ($\alpha =x$, $y$, $z$). }

Despite the seeming similarity of Eqs.\ (\ref{eq3}) and (\ref{eq4}), there
is a significant quantitative difference in the effective fields of the
nuclei acting on the electron spin $A\overrightarrow{F}$ and of the electron 
$\frac{1}{2}A_{j}$ ($S=1/2)$ acting on the nuclei since a large number $N>>1$
of nuclear spins are involved in the QD. The latter property with regard to
the inequality $\omega _{e}>>\omega _{n}$ means a large difference in the
typical precession periods for electron and nuclei spins. Thus, for a
moment, we can consider the time-dependence of $\left\langle \overrightarrow{%
F}\right\rangle $ that is given in a parametric representation. The
corresponding solution of Eq.\ (\ref{eq3}) with respect to $X(t)\equiv
\left\langle S_{X}\right\rangle $, $Y(t)\equiv \left\langle
S_{Y}\right\rangle $, $Z(t)\equiv \left\langle S_{Z}\right\rangle $ and
initial conditions $X(0)=Y(0)=0$, $Z(0)=1/2$ reads 
\begin{equation}
X(t)=\frac{AF_{x}}{2\Omega ^{2}}(AF_{z}+\omega _{e})(1-\cos \Omega t)+\frac{%
AF_{y}}{2\Omega }\sin \Omega t;  \label{eq5}
\end{equation}%
\begin{equation}
Y(t)=\frac{AF_{y}}{2\Omega ^{2}}(AF_{z}+\omega _{e})(1-\cos \Omega t)-\frac{%
AF_{x}}{2\Omega }\sin \Omega t;  \label{eq6}
\end{equation}%
\begin{equation}
Z(t)=\frac{1}{2}-\frac{A^{2}(F^{2}-F_{z}^{2})}{2\Omega ^{2}}(1-\cos \Omega
t).  \label{eq7}
\end{equation}

The frequency of electron spin precession 
\begin{equation}
\Omega =\sqrt{A^{2}F^{2}+2AF_{z}\omega _{e}+\omega _{e}^{2}}  \label{eq8}
\end{equation}%
has the simple physical meaning of the Zeeman frequency of electron in the
total field $\overrightarrow{H}_{e}=\overrightarrow{B}_{e}+A\overrightarrow{F%
}$ composed of external and internal nuclear fields.

Actually, the components of the total nuclear spin moment ${F}_{\alpha }(t)$%
, ($\alpha =x$, $y$, $z$) is affected by the external magnetic field $%
\overrightarrow{B}_{n}$ and the effective field of an electron $%
A\left\langle \overrightarrow{S}\right\rangle $, which is oscillating with a
high frequency $\Omega $ near some mean value $\overline{A\left\langle 
\overrightarrow{S}\right\rangle }$. This mean value also changes slowly with
a typical nuclear frequency $\Omega _{n}\approx \left\vert \overrightarrow{B}%
_{n}\right\vert +A/2$. By taking a small interval in time $\Delta t$ ($%
<<\Omega_{n}^{-1}$) to be longer than the period of electron beats (i.e., $%
\Delta t>>\Omega^{-1}$), one can average over the rapidly oscillating
electron contribution in Eq.\ (\ref{eq4}) and arrive at a system of
equations with respect to ${F}_{\alpha }={F}_{\alpha }(t)$: 
\begin{equation}
\frac{d}{dt}F_{x}=-\omega _{n}F_{y}+A\overline{Y(t)}F_{z}-A\overline{Z(t)}%
F_{y};  \label{eq9}
\end{equation}%
\begin{equation}
\frac{d}{dt}F_{y}=\omega _{n}F_{x}+A\overline{Z(t)}F_{x}-A\overline{X(t)}%
F_{z};  \label{eq10}
\end{equation}%
\begin{equation}
\frac{d}{dt}F_{z}=A\overline{X(t)}F_{y}-A\overline{Y(t)}F_{x}.  \label{eq11}
\end{equation}%
Here the bars mean the procedure of setting $\sin \Omega t$ and $\cos \Omega
t$ to zero when Eqs.\ (\ref{eq5})-(\ref{eq7}) are substituted into Eqs.\ (%
\ref{eq9})-(\ref{eq11}).

Clearly, Eqs.\ (\ref{eq9})-(\ref{eq11}) are non-linear. Fortunately, we
obtain identically zero for the time-derivative $\stackrel{{\bf {\cdot }}}{F}%
_{z}={0}$ that provides conservation of the nuclei spin projection $F_{z}$ on 
the external
magnetic field during the dephasing process described by Eq.\ (\ref%
{eq7}). Thus, the dispersion of $F=\left\vert \overrightarrow{F}\right\vert $
and $F_{z}$, caused by the thermal fluctuation in an ensemble of nuclear
spins associated with QDs, results in the dephasing of electron spin.

Mathematically, evolution of the spin polarization in an ensemble of QDs can be
reduced to the problem of averaging Eq.\ (\ref{eq7}) over the distribution
functions for total nuclear spin [$P(F)$] and for its $z$-projection $\mu
\equiv F_{z}$ [$P_{z}(\mu )$]: 
\begin{equation}
\left\langle Z(t)\right\rangle _{T}=\int \int Z(t)P_{z}(\mu )P(F)d\mu dF.
\label{eq12}
\end{equation}%
The corresponding distribution functions were found in Ref.\ \onlinecite{RybSem}
(see also Ref.\ \onlinecite{Wolff}). In the case of non-saturated nuclear spin
polarization, Eq.\ (\ref{eq12}) can be well approximated by the expression 
\begin{equation}
\left\langle Z(t)\right\rangle _{T}=C\int_{0}^{NI}dF\int_{-F}^{F}d\mu F\exp
(-\frac{F^{2}}{\sigma }-\frac{\omega _{n}\mu }{T})Z(t),  \label{eq13}
\end{equation}%
where $\sigma =\frac{2}{3}I(I+1)N$ and $C$ is a constant. The averaging in
Eq.\ (\ref{eq13}) can be performed numerically for different experimental
situations. Some examples of these calculations are presented in Fig.\ 1 for
the case of relatively high temperature $T\gg A\sigma \omega _{n}/\omega
_{e} $.

Two important conclusions follow immediately from Eqs.\ (\ref{eq7}) and (\ref%
{eq13}) (see also Fig.\ 1). First, electron spin dephasing does not fully
relax its own initial spin polarization $Z_{0}=Z(0)$. The minimal rest $%
Z_{\infty }=\lim_{t\rightarrow \infty }\left\langle Z(t)\right\rangle _{T}$
of $Z_{0}$ corresponds to zero magnetic field: $Z_{\infty }=\frac{1}{3}Z_{0}$%
. In a strong enough magnetic field ($B\rightarrow \infty $), dephasing will
be fully suppressed: $Z_{\infty }\rightarrow Z_{0}$. Intermediate cases can
be traced with high enough accuracy if we substitute high-temperature
approximations for the average values of $\mu $ and $\mu ^{2}$ in Eq.\ (\ref%
{eq13}): $\left\langle \mu \right\rangle _{T}\simeq \frac{1}{3}\left\langle
F^{2}\right\rangle \frac{\omega _{n}}{T}=\frac{\sigma \omega _{n}}{2T}$ and $%
\left\langle \mu ^{2}\right\rangle _{T}\simeq \frac{1}{3}\left\langle
F^{2}\right\rangle =\frac{\sigma }{2}$. A more detailed approximation is of
no interest since it corresponds to saturation of nuclear spin polarization
along electron spin polarization that results in insignificant dephasing.
The final result reads 
\begin{equation}
Z_{\infty }=\frac{\sigma (\frac{1}{2}A^{2}+A\omega _{e}\frac{\omega _{n}}{T}%
)+\omega _{e}^{2}}{\sigma (\frac{3}{2}A^{2}+A\omega _{e}\frac{\omega _{n}}{T}%
)+\omega _{e}^{2}}Z_{0}.  \label{eq14}
\end{equation}

The second conclusion is that dispersion of Zeeman frequency $\Omega $
controls the electron dephasing rate. Hence, instead of calculating $%
\left\langle Z(t)\right\rangle _T$ from Eq.\ (\ref{eq13}), we can find the
distribution function for $\Omega $; then, the dephasing rate can be
estimated as the width of this distribution. In other words, the dephasing
rate can be obtained from the width of electron spin resonance lineshape 
\begin{equation}
g(\omega )=\int \int \delta (\omega -\Omega )P_z(\mu )P(F)d\mu dF.
\label{eq15}
\end{equation}

Using the definitions in Eqs.\ (\ref{eq13}) and (\ref{eq8}) as well as the
new integrand variables $x=F/\sqrt{\sigma }$, $y=\mu /\sqrt{\sigma }$, Eq.\ (%
\ref{eq15}) can be rewritten as 
\begin{equation}
g(\omega )=C^{\prime }\omega e^{-\omega ^{2}\tau _{0}^{2}}\sinh \left[
(2\omega _{e}\tau _{0}+\sqrt{\sigma }\omega _{n}/T)\omega \tau _{0}\right] .
\label{eq17}
\end{equation}%
where $\tau _{0}=1/A\sqrt{\sigma }$. The dephasing rate $\tau _{d}^{-1}$ is
expected to be the half width of the $1/e$ decay in the maximal intensity of 
$g(\omega )$. Equation\ (\ref{eq17}) predicts a weak dependence of $\tau
_{d}^{-1}$ on the parameter $2\omega _{e}\tau _{0}+\sqrt{\sigma }\omega
_{n}/T$, which is proportional to the external magnetic field: 
\begin{equation}
\tau _{d}^{-1}=\varphi (2\omega _{e}\tau _{0}+\sqrt{\sigma }\omega
_{n}/T)\tau _{0}^{-1},  \label{eq18}
\end{equation}%
where the function $\varphi (x)$ falls into the region between (0.69,1) (see
Fig.\ 2). Thus, by the order of magnitude, $\tau _{d}^{-1}\simeq \tau
_{0}^{-1}=A\sqrt{\frac{2}{3}I(I+1)N}$.

At the first glance, it would seem that the efficiency of the phase
relaxation increases with the QD volume $V_{0}$ as $\sqrt{N}=\sqrt{n_{i}V_{0}%
}$, where $n_{i}$ is the concentration of isotopes with the nuclear spin $I$%
. In actuality, however, we should take into account that the constant of
the contact interaction $A$ is proportional to the electron spin density at
a nuclear site.~\cite{HFI} As a result, the rate $\tau _{d}^{-1}$ must
reveal an inverse dependence on $\sqrt{V_{0}}$.

Let us estimate the HFI constant $A$ in terms of ENDOR experimental data in
Si:P.~\cite{Feher} According to the definition, 
\begin{equation}
A=\frac{8\pi }{3}g_{e}g_{n}\mu _{B}\mu _{n}\left\vert \Psi \left( 
\overrightarrow{r}_{n} \right) \right\vert ^{2}\eta .  \label{eq19}
\end{equation}%
Here $g_{e}$ and $g_{n}$ are electron and nuclear g-factors, $\mu _{B}$ and $%
\mu _{n}=\left( m_{e}/m_{p}\right) \mu _{B}$ the Bohr and nuclear magnetons, 
$m_{e}$ and $m_{p}$ the electron and proton masses, $\left\vert \Psi \left( 
\overrightarrow{r}_{n} \right) \right\vert ^{2}$ is the electron 
envelope function density at a nuclear lattice site $\overrightarrow{r}_{n}$,
and the parameter $\eta $ reflects the enhancement of electron density due to 
the Coulomb singularity at the nuclear core [see, for details, 
Ref.~\onlinecite{OptOr} and Eq.~(2.78) therein]. In the case 
of $\mathop{\rm Si}^{29}$, this parameter is $\eta =186$.

To find the parameter $A$ for QD electrons in natural Si, 
we can use the Feher's experimental data~\cite{Feher} obtained for
a shallow donor electron in the presence of $\mathop{\rm Si}^{29}$ separated
by $[400]a$ from the donor centrum ($4a=5.43\times 10^{-8}\mathop{\rm cm}$): 
$A_{400}/2\pi \hbar =7.72\ \mathop{\rm MHz}$. At the same time, the donor 
electron density $\left\vert \Psi \left( \overrightarrow{r}_{400}\right)
\right\vert ^{2}$ calculated at this location
is $ 0.4\times 10^{21}\mathop{\rm cm}^{-3}$.  Then, by utilizing the functional 
form of Eq.~(\ref{eq19}) that is proportional to the square of the electron
density, the HFI constant A at a nuclear site $\overrightarrow{r}_{n}$
can be expressed as $A=A_{400}\left\vert \Psi \left( \overrightarrow{r}_{n}
\right) \right\vert ^{2}/ \left\vert \Psi 
\left( \overrightarrow{r}_{400}\right) \right\vert ^{2} $. 
If a uniform envelope function is assumed inside the QD,
$\Psi \left( \overrightarrow{r}_{n} \right) =1/\sqrt{V_{0}}$, one can
find $A/2\pi \hbar $ = $ 1.93\ \mathop{\rm kHz}$ with the QD diameter and 
thickness of $500\ \mathop{\rm \text{\AA}}$ and $50\ \mathop{\rm \text{\AA}}$, 
respectively.

Then, in terms of the uniform $\Psi $-function approximation, we can estimate
the number of isotopes $\mathop{\rm Si}^{29}$ in a QD: $N$ = $xV_{0}/\Omega
_{0}$ = $2.29\times 10^{4}$, where $x=0.0467$ is abundance of isotopes $%
\mathop{\rm Si}^{29}$ and $\Omega _{0}=20\ \mathop{\rm \text{\AA}}^{3}$ the
volume per one $\mathop{\rm Si}$ atom. Thus, one can find $\sigma $ = $\frac{%
2}{3}I(I+1)N$ = $1.14\times 10^{4}$ and $\tau _{0}$=$0.54\times 10^{-6}$ $%
\mathop{\rm s}$.

Let us also consider the final loss of electron polarization after dephasing
reaches a saturation at $t\gg \tau _{0}$. According to Eq.\ (\ref{eq14}),
the relative loss of initial electron polarization $Z_{0}$ is 
\begin{equation}
\frac{Z_{0}-Z_{\infty }}{Z_{0}}=\frac{1}{3/2+(1+\varepsilon )(\omega
_{e}\tau _{0})^{2}}  \label{eq21}
\end{equation}%
where $\varepsilon =\left( \sigma A/T\right) \left( \omega _{n}/\omega
_{e}\right) $. The value of $\sigma A$ which is independent of electron
radius localization and abundance of isotopes $\mathop{\rm Si}^{29}$
estimates to be approximately 0.025 K. Thus, with the exception of extremely
low temperatures, we expect the parameter $\varepsilon $ to be small.

We can see that Eq.\ (\ref{eq21}) looks similar to that of Hanle effect~\cite%
{OptOr} but with the opposite meaning: increasing magnetic field leads to
enhancement (not suppression) of spin polarization. It reflects the
competition of random field of nuclei, $H_{HF}=\hbar /g_{e}\mu _{B}\tau _{0}$
(that is on the order of 0.1-1 G in the case of Si QDs), and the external
magnetic field $B$. Accordingly, dephasing is not important if $B\gg H_{HF}$%
. It is interesting to note that $H_{HF}$ can reach hundred G in QDs
composed of III-V semiconductors.~\cite{KhaetLoss}

In conclusion, we have considered the influence of an external magnetic
field ${\vec{B}}$ on the electron phase spin relaxation in QDs. The
dispersion of nuclear effective local fields $\overrightarrow{H}_{HF}$ over
the ensemble of the QDs results in electron spin dephasing. The dephasing
rate can be estimated as a spin precession frequency caused by field $%
\overrightarrow{H}_{HF}$, so it depends weakly on an external magnetic field 
${\vec{B}}$ provided nuclear spin polarization (or magnetization) is not
saturated. On the other hand, reduction of electron spin polarization along $%
{\vec{B}}$ during the dephasing process does not lead to full
depolarization. If the magnetic field is suitably strong, $B\gg H_{HF}$, one
can expect an insignificant electron spin depolarization caused by the
dephasing process considered.

The result obtained in this study is directly relevant to quantum computing 
with a large number of qubits (i.e., a large number of QDs containing single 
electrons). Clearly, random variation in the spin precession frequency 
and phase across the qubits is deterimental; for example, it can break/negate
the initial state preparation in a relatively short time frame of 
approximately $\tau _{0}$. Special methods must be applied to compensate 
this distortion.  Our work suggests an effective way to suppress
the dispersion of random local fields, making QDs a more attractable
candidate for quantum computing.

The work performed at North Carolina State University was supported, in
part, by the Office of Naval Research and the Defense Advanced Research
Projects Agency.

\begin{figure}[tbp]
\caption{Spin dephasing of QD electrons calculated for three values of
magnetic fields ($\protect\omega _{e}\protect\tau _{0}=$0, 1, and 2). The
temperature is assumed to be relatively high: i.e., $T\gg A\protect\sigma 
\protect\omega _{n}/\protect\omega _{e}$ (non-saturated nuclear spin
polarization).}
\end{figure}

\begin{figure}[th]
\caption{Calculated value of function $\protect\phi (x)$. }
\end{figure}

%\end{multicols}


\begin{references}
\bibitem{KhaetNaz} A. V. Khaetskii and Y. V. Nazarov, \prb {\bf 64}, 125316
(2001).

\bibitem{Erling} S. I. Erlingsson, Y. V. Nazarov, and V. I. Fal'ko, \prb 
{\bf 64}, 195306 (2001).

\bibitem{OptOr} M. I. Dyakonov and V. I. Perel, in {\em Optical Orientation}%
, edited by F. Mayer and B. P. Zakharchenya (North-Holland, Amsterdam,
1984), pp.\ 11-71.

\bibitem{KhaetLoss} A. V. Khaetskii, D. Loss, and L. Glazman, \prl {\bf 88},
186802 (2002).

\bibitem{RybSem} S. M. Ryabchenko and Yu. G. Semenov, Zh. Eksp. Teor. Fiz. 
{\bf 84}, 1419 (1983) [Sov. Phys. JETP {\bf 57}, 1301 (1983)].

\bibitem{Durst} A. C. Durst, R. N. Bhatt, and P. A. Wolff, \prb {\bf 65},
235205 (2002).

\bibitem{Wolff} See: P. A. Wolff, in {\em Diluted Magnetic Semiconductors},
edited by J. K. Furdyna and J. Kossut (Academic, Boston, 1988), Vol. 25 of
Semiconductors and Semimetals.

\bibitem{HFI} {\em Hyperfine Interaction}, edited by A. J. Freeman and R. B.
Frankel (Academic, New York, 1967).

\bibitem{Feher} G. Feher, Phys. Rev. {\bf 114}, 1219 (1959).
\end{references}
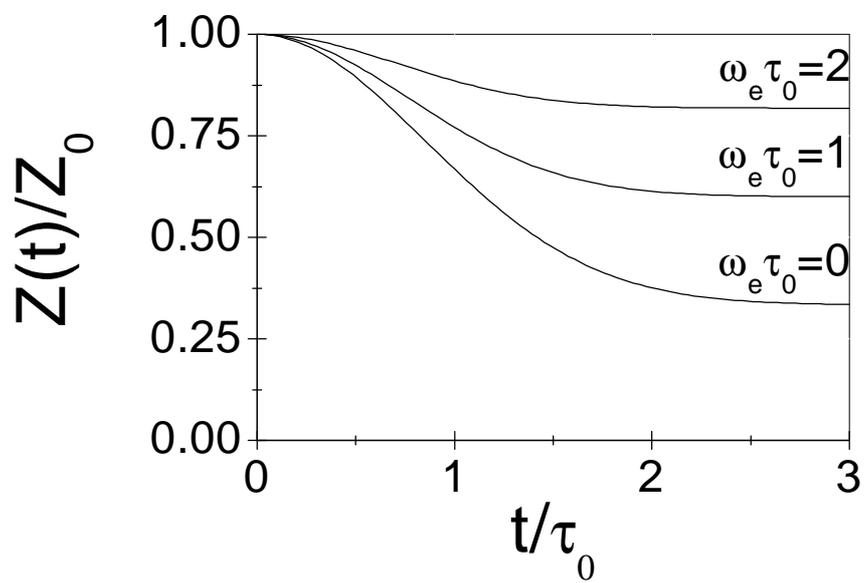
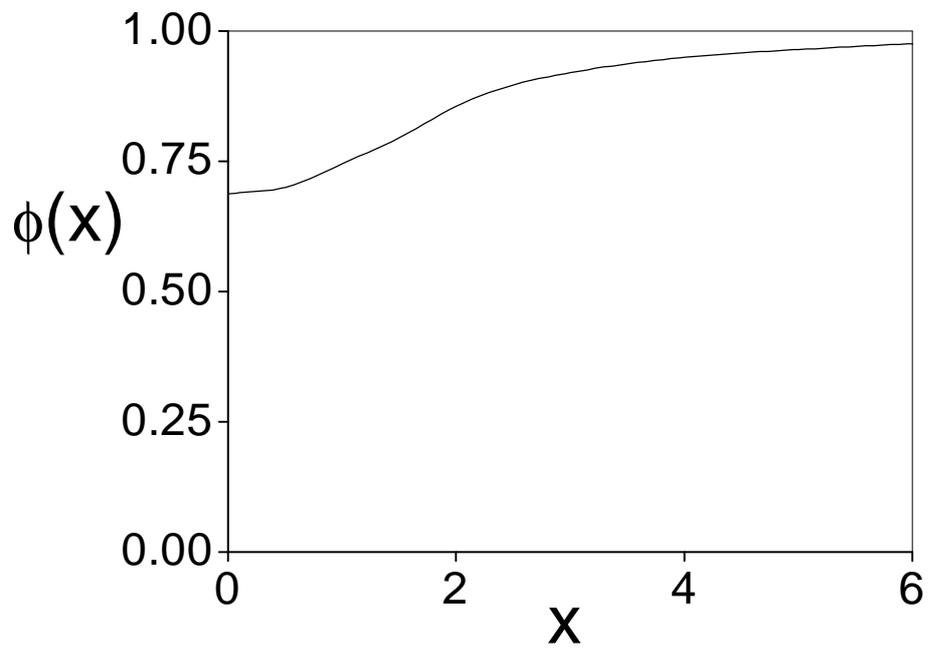
\end{document}